\documentclass[12pt]{article}

\makeatletter
\renewcommand{\ps@plain}{%
        \renewcommand{\@evenhead}{}%
        \renewcommand{\@oddhead}{}%
        \renewcommand{\@evenfoot}{\hfil\small{\textbf\thepage}\hfil}%
        \renewcommand{\@oddfoot}{\@evenfoot}}%

\renewcommand\section{\@startsection {section}{1}{\z@}%
                                   {-3.5ex \@plus -1ex \@minus -.2ex}%
                                   {2.3ex \@plus.2ex}%
                                   {\reset@font\bfseries}}
\renewcommand\subsection{\@startsection{subsection}{2}{\z@}%
                                     {-3.25ex\@plus -1ex \@minus -.2ex}%
                                     {1.5ex \@plus .2ex}%
                                     {\reset@font\itshape}}

\newlength{\capsize}
\renewcommand{\@makecaption}[2]{%
  \vskip\abovecaptionskip
  \sbox\@tempboxa{\small{\bfseries #1}\/: #2}%
  \ifdim \wd\@tempboxa >\capsize
   {\advance\leftskip by 0.1\textwidth \advance\rightskip by 0.1\textwidth
    {\small {\bfseries #1}: #2}\par}
  \else
    \hbox to\hsize{\hfil\box\@tempboxa\hfil}%
  \fi
  \vskip\belowcaptionskip}

\makeatother
\usepackage{amssymb}
\usepackage{latexsym}
\usepackage{graphics}
\usepackage{graphicx}
\usepackage{psfrag}

\title{\large \bf Loop Quantum Gravity and Asymptotically Flat Spaces}

\author{{\normalsize \bf Matthias
Arnsdorf}\thanks{m.arnsdorf@ic.ac.uk},
\\ 
\textit{\normalsize Blackett Laboratory, Imperial College of Science 
Technology and Medicine,}
\\ 
\textit{\normalsize London SW7 2BZ, United Kingdom.}
 }

\date{{\normalsize \bf August 15, 2000}\\ {\normalsize Based on talk
presented at
the} \\{ \normalsize 9th Marcel Grossmann meeting in Rome, July 2000}}

\newcommand{\Ho}{\mathcal{H}_{\omega}} 
\newcommand{\Hil}{\mathcal{H}}
\newcommand{\A}{\mathfrak{A}}
\newcommand{\I}{\mathfrak{I}} 

\newcommand{\unit}{\mathbb{I}}

\newcommand{\ip}[2]{\left\langle #1 , #2 \right\rangle}
\newcommand{\norm}[1]{\left\| #1 \right\|} 

\newcommand{\bra}[1]{\langle #1|}
\newcommand{\ket}[1]{|#1 \rangle}

\newcommand{\om}{\omega}

\newcommand{\ba}{{\mathbf a}}
\newcommand{\bb}{{\mathbf b}}

\newcommand{\Aia}{A^{i}_{a}}

\newcommand{\g}[1][]{\gamma_{#1}}

\newcommand{\Aob}{\mathfrak{B}_\mathrm{aux}}

\newcommand{\cyl}{\mathfrak{C}} 
 
\begin{document}

%Hilbert spaces 
\newcommand{\Ha}{\mathcal{H}_\mathrm{aux}} 
\newcommand{\Hw}{\mathcal{H}_W} 
\newcommand{\Hg}{\mathcal{H}_{\G}} 
\newcommand{\Hk}{\mathcal{H}_\mathrm{kin}} 
\newcommand{\Hb}{\mathcal{H}_{\B}} 
\newcommand{\Hq}{\mathcal{H}_{\Q}} 
\newcommand{\Hbk}{\Hb^\mathrm{kin}} 
\newcommand{\Hqk}{\Hq^\mathrm{kin}} 

\newcommand{\ra}{\rightarrow}

\newcommand{\snorm}[1]{\left\| #1 \right\|_{\infty}}

\newcommand{\R}{\mathbb{R}}
\newcommand{\C}{\mathbb{C}}

%weave 
\newcommand{\gw}{\Gamma_{W}} 
\newcommand{\gb}{\Gamma_{\B}} 
\newcommand{\W}{\mathcal{W}} 
\newcommand{\B}{\mathcal{B}} 
\newcommand{\BR}{\B|_R} 
\newcommand{\QR}{\Q|_R} 
 
\newcommand{\hBR}{\hat{\B}|_R} 
\newcommand{\hQR}{\hat{\Q}|_R} 
\newcommand{\Q}{\mathcal{Q}} 
\newcommand{\gq}{\Gamma_\Q} 
 
%representations 
\newcommand{\po}{\pi_{\om}} 

\maketitle

\begin{abstract}

After motivating why the study of asymptotically flat spaces is
important in loop quantum gravity, we review the extension of the
standard framework of this theory to the asymptotically flat sector
detailed in~\cite{arnsdorf99b}.  In particular, we provide a general
procedure for constructing new Hilbert spaces for loop quantum gravity
on non-compact spatial manifolds. States in these Hilbert spaces can
be interpreted as describing fluctuations around fiducial fixed
backgrounds.  When the backgrounds are chosen to approximate classical
asymptotically flat 3-geometries this gives a natural framework in
which to discuss physical applications of loop quantum gravity,
especially its semi-classical limit.  We present three general
proposals for the construction of suitable backgrounds, including one
approach that can lead to quantum gravity on anti-DeSitter space as
described by the Chern-Simons state.

\end{abstract}

\section{Motivating Asymptotic Flatness}

Remarkable progress has been made in the field of non-perturbative
(loop) quantum gravity in the last decade or so and it is now a
rigorously defined kinematical theory.  We are at the stage where
physical applications of loop quantum gravity can be studied and used
to provide checks for the consistency of the quantisation
programme. Equally, old fundamental problems of canonical quantum
gravity such as the problem of time or the interpretation of quantum
cosmology need to be reevaluated seriously.  The purpose of this
report is to suggest that all these issues can be tackled most
profitably in the asymptotically flat sector of quantum gravity and to
discuss the extension of loop quantum gravity to this
sector that was developed in~\cite{arnsdorf99b}. 

\subsection{Canonical quantum gravity}

One of the defining features of general relativity is its
active diffeomorphism invariance. 
This implies that we
can take all fields on the space-time manifolds and reassign their
values to different points without changing predictions. Hence,
points in the space-time have no significance of there own,
or in other words there is no fixed reference frame.
This leads to the relational nature of observables
as is exemplified in Einstein's hole argument: values of fields are
only physically meaningful (observable) if they are given with
reference to points defined by the values of other fields.

Physics in this context is difficult in practice. But usually,
to study concrete gravitational systems and make predictions we do not
need such generality. One can break the diffeomorphism invariance and
introduce reference frames or introduce extra symmetries. A very
convenient way to do this  is to study the asymptotically
flat sector of general relativity, as is discussed below.

Another set of reasons to study asymptotically flat spaces, originates
in the structure of quantum theory. General relativity deals naturally
with the physics of the entire universe, which is the definitive
closed system. Quantum mechanics on the other hand is plagued with
interpretational difficulties when attempting to describe genuine
closed systems. More specifically, we need an external observer to
make sense of predictions made for a given physical system.

For these reasons it is unlikely that we will succeed in a complete
and comprehensive theory of quantum gravity, without major changes
to the ingredient theories viz.\ general relativity and quantum
mechanics. This motivates us to study  sectors of the theory, where we
can make progress in addressing physical problems, while avoiding many
of the foundational concerns.

\subsection{Asymptotically flat spaces}

Many physical applications of general relativity involve the study of
asymptotically flat spaces. Here one restricts attention
to the subset of the phase space in which the variables approach the
flat configuration at spatial infinity. Physically, this means that we
are breaking diffeomorphism invariance at infinity and introducing a
reference frame there. We can interpret this as the idealised
description of an isolated gravitational system viewed by an observer
in the ambient environment. This brings several technical advantages:

\begin{enumerate}
\item Since space is no longer compact we need to add boundary terms
to the Hamiltonian, which are integrals over a two sphere at
infinity. It follows that the Hamiltonian no longer vanishes but has
the value of the ADM-energy: the total energy of the isolated
gravitational region.

\item In addition, we recover momentum and angular momentum observables
at infinity. Together these functions generate the Poincar\'e group
at infinity, which is the is the symmetry group of the isolated
region.

\item In particular, we can also introduce a notion of asymptotic
time translations generated by the energy observable. This now makes
sense since we can interpret this time variable with respect to the
fixed reference frame at infinity

\item The introduction of an external environment also allows a
natural interpretation of a corresponding quantum theory in terms of
observers at infinity. The setup essentially amounts to the study of a
gravitational system in a box, which resembles the standard
applications of quantum mechanics.
\end{enumerate}

For these reasons the study of asymptotically flat spaces provides a
very promising testing ground for any theory of quantum
gravity. Indeed, it is likely that we should obtain a quantum theory
for this special case even if it is not possible to quantise full general
relativity. In addition, many of the physical applications
of a quantum gravity theory fall into this sector.

\begin{enumerate}

\item The study of the weak field limit of quantum gravity is a
crucial test for quantum gravity. In the absence of concrete
experimental evidence consistency with results of perturbative quantum
gravity is essential. This involves the investigation of perturbations
of flat space and the calculation of scattering amplitudes of
gravitons.

\item Black Holes and more general isolated gravitational systems provide
another fascinating application for quantum gravity. Again we need to
recover results from quantum theory of curved spaces, which are
theoretically very robust.

\item A genuine experimental test for quantum general relativity might
be possible with the study of $\gamma$-ray bursts. The idea is to
detect any breaking of Poincar\'e invariance by exploiting the
accumulation of tiny effects over very long distances.

\end{enumerate}

\subsection{Loop quantum gravity and the GNS construction}

Loop quantum gravity is a quantum theory of the 3-geometry of a
spatial manifold $\Sigma$. In contrast to the geometrodynamical
approach to canonical quantum gravity, the phase space coordinates in
loop quantum gravity are given by connection and triad fields defined
on $\Sigma$. A well-defined kinematical quantum framework for this
theory has been developed, and several proposals for incorporating
dynamics or the Hamiltonian constraint into the theory are currently under
study.

In this framework it turns out that fundamental excitations of
geometry have 1-dimensional support. More precisely, excitations are
 concentrated on graphs embedded in $\Sigma$ providing a polymer like
picture of space-time at the Planck scale. 

The standard approach to
loop quantum gravity is only applicable to the case where the
3-manifold $\Sigma$ is compact. Roughly, this can be seen as
follows. To describe  a geometry on $\Sigma$ with a quantum state we
need excitations of geometry in every macroscopic region of
$\Sigma$. Since the excitations are concentrated on embedded graphs
this entails that underlying the definition of our state should be a
graph that spreads through all of $\Sigma$, i.e.\ every macroscopic
region of $\Sigma$ should contain vertices and edges of the graph.

For non-compact $\Sigma$ this implies that we need graphs with an
infinite number of vertices and edges and it turns out that states
based on such graphs are not included in the standard Hilbert space.

The solution we review in this report (c.f.~\cite{arnsdorf99b}) is
motivated by an analogy with thermal field theory (TFT). Here a
similar problem arises when considering fields at a finite
temperature. These are described by a condensate of an infinite number
of photons, which cannot be described within the standard Fock space.
The issue that needs to be addressed in both cases is how to describe
an infinite number of excitations.

To do this we take an algebraic approach to quantum
theory. This means that we regard algebras of observables as the
primary objects, with states just arising as elements on the spaces on
which we choose to represent the algebra. Evolution becomes an
automorphism of the observable algebra and all physical questions then
involve the calculation of expectation values of the relevant
operators. This view point gives us the freedom to change between
representations according to which sector of the quantum theory we are
interested in.

We will show how the Gel'fand Naimark Segal (GNS) construction
provides us with representations that can be interpreted naturally as
describing the excitations of fiducial fixed background states. In
particular, these background states can be chosen to describe
asymptotically flat
geometries on non-compact 3-manifolds.

This will be investigated in the last part of this report where we
discuss three possible frameworks for the  construction of  suitable
backgrounds. Specific choices of these backgrounds can be interpreted as
semi-classical states, also referred to as vacua of loop quantum
gravity, and their excitations should describe field theory on curved
spaces.

\section{An observable algebra for loop quantum gravity}

In this section we review the standard kinematical framework of loop
quantum gravity. 
In particular, we specify the algebra $\Aob$ of elementary classical
observables that we wish to represent. We will take a brief look at
how the standard representation of this algebra arises when space is
compact before looking at new
representations constructed via the GNS construction in the following sections.
 
The classical phase space for general relativity in the connection
variables~\cite{ashtekar86} is the co-tangent bundle of the
configuration space, given by $\mathfrak{su}(2)$ valued connection
one-forms with co-ordinates\footnote{Here $a$ denotes a spatial index
and $i$ a Lie-algebra index.}  $A^i_a$ on a spatial 3-manifold
$\Sigma$. The conjugate variable is a desitised triad $\tilde{E}^a_i$,
which takes values in the dual of the Lie algebra $\mathfrak{su}(2)$.
These triads can be considered as duals to two forms $e_{abi} \equiv
\epsilon_{abc}\tilde{E}^{c}_i$. The dynamics of general relativity on
spatially compact manifolds is then completely described by the Gauss
constraints which generate SU(2)-gauge transformations, the
diffeomorphism constraints which generate spatial diffeomorphisms on
$\Sigma$, and the Hamiltonian constraint, which is the generator of
coordinate time evolution.
 
Quantisation proceeds in two steps. First we seek a representation of
the algebra of classical variables $\Aob$ on some auxiliary Hilbert
space $\Ha$. The second step is to obtain operator versions of the
classical constraints and to then impose these on the Hilbert space to
obtain a reduced space of physical states along with a representation
of the subalgebra of observables that commute with the constraints.
In this review we will restrict ourselves to the kinematical sector
and focus on the first step of this procedure, more details can be
found in~\cite{arnsdorf99b}.

To obtain the classical algebra of elementary functions which can be 
implemented in the quantum theory, we need to integrate the canonically 
conjugate variables, $\Aia$ and $e_{iab}$, against suitable smearing 
fields. In usual quantum field theory, these test fields are 
three-dimensional. However, in canonical quantum general relativity, due to 
the absence of a background metric, it is more convenient to smear $n$-forms 
against $n$-dimensional surfaces~\cite{rovelli90,ashtekar92,ashtekar98}. 
 
Configuration observables (depending only on the connection) can be
constructed through holonomies of connections.  Given an embedded
graph $\Gamma$ which is a collection of  paths
$\{\g[1],\ldots,\g[n]\} \in \Sigma$, and a smooth function $f$ from
$\mathrm{SU(2)}^n$ to $\mathbb{C}$, we can construct cylindrical
functions of the connection:
\[ 
    \psi_{f,\Gamma}(A) = f(H(A,\g[1]),\ldots,H(A,\g[n])). 
\] 
$H(A,\g[i]) \in \mathrm{SU(2)}$ is the holonomy assigned to the edge
$\g[i]$ of $\Gamma$ by the connection $A$. We denote by $\cyl$, the
algebra generated by all the functions of this form. To obtain
momentum variables, we smear the two-forms $e_{abi}$ against
distributional test fields $t^i$ which take values in the dual of
$\mathfrak{su}(2)$ and have two dimensional support. This gives us
\[ 
    E_{t,S} \equiv \int_{S} e_{abi} t^i dS^{ab}, 
\] 
where $S$ is a two-dimensional surface embedded in $\Sigma$. More 
precisely~(c.f.~\cite{ashtekar98}), we require that \mbox{$S = \bar{S} - 
\partial\bar{S}$}, where $\bar{S}$ is any \emph{compact}, analytic, two 
dimensional submanifold of $\Sigma$. 
 
The elements of $\cyl$ and the functions $E_{t,S}$ are the variables
that we wish to promote to quantum operators. They form a large enough
subset of all classical observables in the sense that they suffice to
distinguish phase space points.  The algebra of elementary
observables, $\Aob$, is the algebra generated by the cylindrical
functions and vector fields on $\cyl$ associated to the momentum
variables, details can be found in~\cite{ashtekar98}.
 
\subsection{The standard representation}

We now describe briefly how $\Ha$ with its representation of $\Aob$ is
constructed. 
The Hilbert space $\Ha$ is chosen to be the completion of the space of
cylindrical functions $\cyl$, first in the sup norm:
\[ 
    \snorm{\psi_{f,\Gamma}} = \sup |\psi(A)_{f,\Gamma}|, 
\] 
and then in the norm based on the (\underline{s}tandard) inner product~\cite{ashtekar93}:
\begin{eqnarray}  
    \ip{\psi_1}{\psi_2}_s &=& \int d\mu(A) \psi_1^*(A)\psi_2(A)
\nonumber \\
& \equiv &
    \int_{\mathrm{SU(2)}^n}f^*_1(g_1,\ldots,g_n)f_2(g_1,\ldots,g_n) dg_1\cdots 
    dg_n, \label{def-ip}
\end{eqnarray} 
where $g_i \in \mathrm{SU}(2)$ and $dg$ is the Haar measure on SU(2).
Here, $\psi_1$ and $\psi_2$ are defined with respect to the same
graph. This does not present a loss of generality since given two
states supported on different graphs $\Gamma_1$ and $\Gamma_2$. we can
view them as being both defined on $\Gamma_1 \cap \Gamma_2$, with the
states having trivial dependence on the edges that do not belong to
their respective original graphs.
$\Ha$ can also be 
regarded as space of square integrable functions defined with respect to a 
genuine measure on some completion $\bar{\A}$ of $\A$ as is done 
in~\cite{ashtekar95}.

This Hilbert space carries a multiplicative representation of the
configuration variables $\cyl$ and we are left with the task of
representing the momentum variables. This done by constructing
essentially self adjoint operators $\hat{E}_{t,S}$ on $\cyl$ which can
be extended to $\Ha$. These operators are derivations on $\cyl$ i.e.\
linear maps satisfying the Leibnitz rule, which act on functions
$\psi_{f,\Gamma}\in \cyl$ only at points where $\Gamma$ intersects the
oriented surface $S$. The precise definition of these operators is not
needed for our purposes, but it can be found, e.g.,
in~\cite{ashtekar98}. This choice of operators gives the correct
representation of the classical algebra $\Aob$, which provides us with
a kinematical framework for canonical quantum gravity. In the
following this representation will be referred to as the standard
representation $\pi_s$.

\subsection{Problems with describing non-compact geometries}

Are main interest lies in studying the asymptotically flat sector of
loop quantum gravity and hence we need to have states describing
geometries on non-compact spatial manifolds. In the following we
describe briefly why this is generically not possible in the standard
framework.  As mentioned in the introduction we need cylindrical
functions based on graphs with an infinite number of edges and vertices
to describe excitations everywhere on a non-compact space. Hence, one
first needs to specify what one means by a cylindrical function based
on such an infinite graph. A natural approach is to consider sequences
of cylindrical functions based on larger and larger but finite
graphs. One then finds that in general these sequences are not Cauchy
in the available norms on the Hilbert space $\Ha$ and hence they do
not lie in the standard state space.

\section{Representations induced by background states}

As the standard framework of loop quantum gravity is not applicable to
the asymptotically flat sector we look at how we can construct new
representations of the observable algebra of loop quantum gravity. The
main aim will be to obtain quantum theories that describe excitations
of asymptotically flat backgrounds. In this section we will look at how
this can be done in general using the GNS construction before applying
these techniques to loop quantum gravity.

When faced with an infinite dimensional algebra of observables that we
wish to quantise,  we can no longer rely
on the Stone - Von Neumann theorem to guarantee us the existence of a
unique irreducible representation. It is in fact well-known that in
quantum 
field theory there are an infinite number of representations to choose
from. The obvious questions that arise are how to construct these
representations, how to  select an appropriate
representation for the physical situation at hand and how to interpret
physically the sates that arise in this representation. As we will
motivate in this section, there is a satisfactory answer to these
questions if we are given a suitable background state as is defined
more explicitly below.

To motivate the definition that follows in the next section, we
imagine that we are given a representation $\pi$ of an algebra $\A$ on
a Hilbert space $\Hil$. In addition we require the existence of a
preferred ``vacuum'' state $\Psi$ that is cyclical, which means that the set of
states $\{\pi(\ba)\Psi | \ba \in \A\}$ is dense in $\Hil$. The first
immediate observation is that essentially all states can be identified
with an algebra element, i.e.\ $\ba \leftrightarrow \phi$ if $\phi =
\pi(\ba)\Psi$. We say that $\phi$ is an excitation of the vacuum
corresponding to $\ba$. This should be seen in close analogy to the
construction of the Fock space by repeated action of creation
operators on the vacuum. 

The next observation follows from the introduction of the positive
linear from $\om$ which assigns to every algebra element its vacuum
expectation value (VEV): $\om(\ba) = \ip{\Psi}{\pi(\ba)\Psi}$. Once
$\om$ is specified we can express the inner product between
essentially any two states $\phi_1$, $\phi_2$ as $\ip{\phi_1}{\phi_2} =
\om(\ba^*_1\ba_2)$, where $\pi(\ba_{1/2})\Psi =
\phi_{1/2}$. So in summary, given $\om$ we can answer any physical question
about the system of interest.

The problem faced in field theory is that we are not given a
vacuum as a state in some Hilbert space. In general, all we have
initially is the algebra $\A$ of observables. The question thus is whether we
can reverse the above and construct a representation of $\A$ just
given a positive linear form $\om$ on $\A$, which should be
interpreted as giving the VEV's of all observables. States arising in
this representation should be thought of as excitations of the vacuum
corresponding to $\om$. The answer is in the affirmative and the
procedure is given by the GNS construction detailed below.

\subsection{The GNS construction} \label{gns}

The GNS construction (see, e.g.,~\cite{haag93,landsman87,kadison83} for more 
detailed expositions) allows us to construct a representation of any 
$^*$-algebra $\A$ for any given positive linear form (also called a state) 
$\om$ on this algebra. This is done in three steps: 
\begin{enumerate} 
    \item Using $\om$, define a scalar product on $\A$, regarded as a
linear space over $\mathbb{C}$, by \[ \ip{\ba_1}{\ba_2} = \om(\ba^*_1\ba_2),
\] for $\ba_1,\ba_2 \in \A$. The positivity of $\om$ implies
$\ip{\ba}{\ba} \geq 0$. Hence, we are interpreting the elements of the
algebra directly as excitations of the vacuum, as motivated above.
 
    \item To obtain a positive definite scalar product, we construct
the quotient $\A/\I$ of $\A$ by the null space $\I = \{\ba \in \A |
\om(\ba^*\ba) = 0\}$. We denote the equivalence classes in $\A/\I$ by
$[\ba]$ and we have: \[ \ip{[\ba]}{[\ba]} \equiv \norm{\ba}^2 > 0 \] The
completion of $\A/\I$ in the above norm is the carrier Hilbert space
$\Ho$ for our representation.
 
    \item Finally it can be shown that a representation $\pi_{\om}$ of
$\A$ on $\A/\I$ (which, if $\A$ is a Banach $^*$-algebra, can be
extended continuously to $\Ho$) is given by: \[ \pi_{\om}(\ba)\Psi =
[\ba\bb], \] for $\Psi = \bb \in \Ho$ and $\ba \in \A$.
\end{enumerate} 

Hence we have a procedure at our disposal that provides
representations of observable algebras describing fluctuations around
fiducial backgrounds. Note crucially that this ``vacuum'' is
\emph{not} defined via a state in some Hilbert space. Indeed, the
background need not be realised at all as an element of
$\Ho$. Although, if $\A$ has a unit $\unit$ then the background
corresponds to the cyclic state $[\unit]$.

\section{New representations for loop quantum gravity}

We will now use this technique to construct new representations of
loop quantum gravity. The main technical input
will be to construct positive linear forms on the algebra of
observables that can be interpreted as giving vacuum expectation
values of some background.

\subsection{Cylindrical function backgrounds}

We consider first the most straightforward extension of the standard
loop quantum gravity framework. 
As we have seen the problem there is that sequences of cylindrical
functions based on increasingly large graphs are not physical states.
The crucial point to this subsection is
that these sequences can nevertheless be used to provide us with a
definition of a positive linear form on the algebra of observables.

In the following we will look at this construction in more
detail. Since this will depend on the precise choice of
sequence that one is interested in, we will for simplicity restrict
our attention to a specific class of states.  Let us assume that we
are given an infinite graph $\gq$, i.e. a graph that extends throughout
a non-compact spatial manifold $\Sigma$ but that has (at most) a finite number
of edges and vertices in any compact subregion of $\Sigma$. Associated
to each edge $e_i$ of $\gq$ there should be a normalised cylindrical
function $q_i$ that is a function of the holonomy along $e_i$. This
allows to define the states:
\[
\Q_n = \prod_{i=1}^n q_i,
\]
for any $n$.  As described of above we are interested in the limit $\Q
= \Q_{\infty}$ of this sequence as $n$ tends to infinity which in
general is not an element on $\Ha$.  Nevertheless, we can define the
action of an element of $\Aob$ on $\Q$. The crucial point is that the
elementary quantum observables --- the elements of $\cyl$ and the
derivations on $\cyl$ --- have support on a compact spatial region,
which is a direct consequence of the smearing needed to make sense of
the classical expressions. Hence given an arbitrary
element $\ba \in \Aob$ we proceed as follows:
 
\begin{enumerate} 
    \item Denote the closure of the support of $\ba$ by $R \subset \Sigma$. 

    \item Construct the graph $\gq|_R$ of $\gq$ restricted to $R$: 
    \[ 
         \gq|_R \equiv \bigcup_{e_i \cap R \neq \emptyset} e_i. 
    \]  
In other words, consider the union of all edges which have a non-zero 
intersection with the support of $\ba$. This graph is finite, since $R$ 
is compact and we obtain the state $\QR \in \Ha$ which is given by restricting 
$\Q$ to the graph $\gq|_R$: 
\[ 
     \QR \equiv \mathbb{I} \cdot \prod_{e_i \in \gq|_R} q_i, 
\] 
where $\mathbb{I}(A) = 1$ for all $A$ is the identity function.
This state has unit norm in $\Ha$ since all the $q_i$'s are normalised. 
 
    \item Since $\QR \in \Ha$, the action of $\ba$ on $\QR$ 
    denoted by $\pi(\ba)_s\QR$ is well-defined. It is understood 
    that the region $R$ will depend on $\ba$. 
\end{enumerate} 
 
This allows us to define the positive linear form $\om_{\Q}(\ba)$: 
\begin{equation}\label{form} 
    \om_{\Q}(\ba) = \ip{\Q|_R}{\pi_s(\ba)\Q|_R}_s.
\end{equation} 
 This is well-defined since $\QR$ is an
element of $\Ha$ for all $\ba$. It follows from the fact that
equation~(\ref{def-ip}) defines a true inner product
$\ip{\cdot}{\cdot}_s$ on $\Ha$ that $\om$ is indeed a positive (not
necessarily strictly positive) linear form on $\Aob$.

Given $\om_{\Q}$, we can  proceed with steps 2 
and 3 of the GNS construction outlined in section~\ref{gns} to 
obtain a representation of the algebra $\Aob$. 
It turns out, as described in detail in~\cite{arnsdorf99b}, that the
representation is equivalent to a very intuitive representation
$\pi_{\Q}$ on $\Ha$:
 \begin{equation}\label{def-rep2} 
    \pi_{\Q}(\ba)\psi = \QR^{-1} \pi_s(\ba) (\QR\psi),
 \end{equation}
where 
    $\ba\in\Aob$, $\psi\in\Ha$ and $\QR^{-1}$ denotes the inverse 
    function\footnote{At this 
    point we note an additional requirement for the background state: 
    \mbox{$\Q(A) \neq 0$} for all $A$.  This invertability property is 
    motivated physically since our background state is meant to 
    represent an infinite `condensate of gravitons'. We should be able 
    to annihilate as well as create these gravitons, which motivates 
    invertability.}
  of $\QR$ i.e., $\QR^{-1} \QR = \mathbb{I}$.

 Intuitively, the above representation has a clear
interpretation.  We can regard the algebra of cylindrical functions $\cyl$
as creating and annihilating excitations on the background state. More
general operators then act on this excited ``vacuum''. 
Hence, we have constructed a Hilbert space and representation of
observables on it that describes fluctuations restricted to
essentially compact regions around some fixed infinite background
state.
Note that this representation is truly inequivalent to the standard
one. Roughly, since $\Q|_R$ depends on the algebra element $\ba$,
equation~(\ref{def-rep2}) does not define a unitary map.

The construction we have presented is very general and can be applied
to a large class of background states.
The advantages of this approach are that the final formalism is very
simple. One can use the same (separable) Hilbert space as in the standard
representation and in particular the reduction by the constraints can
be carried out as in the standard approach.
To study quantum gravity on semi-classical, asymptotically flat
geometries one need states that approximate phase space points. 
 There is now a variety of such cylindrical functions available that can be
used in the above approach,
c.f.~\cite{arnsdorf99b,thiemann00,corichi00}.

\subsection{Mixed backgrounds}

Here we look at a possible improvement of the above approach. A
difficulty in studying and interpreting the classical limit of loop
quantum gravity is the fact that states are supported on graphs or
more physically that we have a quantum picture of polymer like
excitations of geometry. The familiar continuum picture has to be
recovered from the study of coarse grained observables. In particular,
we would like to approximate classical values of observable functions
at a particular phase space point to increasing accuracy with the
expectation values of corresponding observable operators in some
semi-classical state as $\hbar \ra 0$. These conditions are not enough
to specify a unique state. In particular, the graph on which the
semi-classical state is to be based is left largely undetermined. This
is due to the fact that in studying the classical limit we are using
operators that are too coarse grained to determine the micro structure
of the quantum state completely.

This suggests naturally that we should really be considering a
statistical mixture of states, which is in fact analogous to what is
done in thermal field theory, where the vacuum state is only specified
by the macroscopic temperature. 

Hence, the
``gravitational vacuum'' is composed of many subsystems, each
described by their own
 micro-state $\ket{\phi_i}$. Given the set of macroscopic variables
that characterise the vacuum let us denote the probability that
a particular subsystem will be in the state $\ket{\phi_i}$ by $P(\phi_i)$. 
The gravitational
vacuum is then given by the density matrix:
\[
\rho = \sum_i P(\phi_i) \ket{\phi_i} \bra{\phi_i},
\]
where $\ket{\phi_i} \bra{\phi_i}$ denotes the projector onto the micro-state
$\ket{\phi_i}$. This can be used to define a positive linear
functional on the algebra of observables:
\[
\om(\ba) \equiv Tr[\rho \ba],
\]
which gives rise to the desired quantum theory.
Again we are especially interested in the case that the density matrix
describes asymptotically flat geometries. In this case the micro-states
$\phi_i$ will be based on infinite graphs and we need to make use of
the techniques of the preceding section to make sense of the above
linear form.

In the case that the macroscopic observable characterising the
gravitational background is the volume of regions of the spatial
manifold $\Sigma$ there is a natural construction to implement the
above based on random lattices,
c.f.~\cite{thiemann00c,bombelli00}. This gives us a mixture of states
based on a large class of graphs.

\subsection{Chern Simons backgrounds}

The preceding constructions depend on the ability to approximate
classical phase space points with cylindrical function states. While,
progress has been made in this direction a major unaddressed issue is
still the dynamics. No semi-classical cylindrical function state has
been proposed so far that solves all the constraints, which should be
a necessary criterion for any true physical state. 

The power of the GNS construction proposed here is that one is not
tied to using cylindrical functions to define suitable approximations
of 3-geometries. As we have seen all that is needed is the definition
of a positive linear form on the observable algebra, which can be
interpreted as giving the vacuum expectation values in some preferred
state.  This section will be devoted to the study of one such
alternative based on the Chern Simons state. While more heuristic at
present we believe that the following approach has many promising
features.

The Chern Simons state was discovered early on~\cite{kodama90} as an exact
solution to all the constraints of quantum general relativity. The
term ``state'' here is used here in a heuristic sense, as it is not an
element of a known Hilbert space. Rather, it is defined as a function on
the classical configuration space, the space of connections:
\[
\Psi_{CS}(A) \equiv \exp\left(- \frac{6}{\Lambda} \int_{\Sigma} Tr[A \wedge dA
+ \frac{2}{3} A \wedge A \wedge A]\right), 
\]
where $\Lambda$ is the cosmological constant.
This state has generated a lot of interest as it has a well-defined
classical limit corresponding to anti-DeSitter
space~\cite{kodama90,smolin95}.

Study of this state has been mainly within the loop representation of
loop quantum gravity, which is in a precise sense dual to the one
described in this report.  Here states are functions of loops or, more
generally, graphs embedded in $\Sigma$ instead of connections. Given a
cylindrical function $\Psi_{f,\gamma}$ based (for simplicity,
generalisations to graphs are straightforward) on a loop $\gamma$ we
can transform to the following state in the loop representation:
\[
\tilde{\Psi}(\gamma) = \int Tr[H(\gamma,A)]\Psi_{f,\gamma}(A) d\mu(A), 
\]
using the cylindrical function measure defined in eq.~(\ref{def-ip}).
One can attempt to do the same with the Chern-Simons state and define:
\begin{equation} \label{CS-transform}
\tilde{\Psi}_{CS}(\gamma) = \int  Tr[H(\gamma,A)]\Psi_{CS}(A) dA
\end{equation}
Since no appropriate measure on the space of connections is known the
above is only heuristic. But crucially, one can nevertheless define
the integral by using Witten's celebrated result that the above
transformation defines knot invariants of \emph{any} embedded loop in
$\Sigma$ (see e.g.~\cite{pullin93} for an overview). This implies that
if we want to interpret the Chern-Simons state as a cylindrical
function then it would have support on all possible loops and graphs
in $\Sigma$.

This is appealing but also leads directly to the fact that the Chern
Simons state is not a physical state of loop quantum gravity as it
cannot be normalised with respect to any known inner product.  The key
to the remainder of this section will again be the fact that we can
still use the Chern-Simons state to define a positive linear function
on the algebra of observables and thus define a quantum theory
describing fluctuations of the Chern-Simons state, which following
arguments in~\cite{smolin95} should correspond to field theory on
Anti-DeSitter space in the semi-classical limit.

So to proceed we propose to define $\om$ via:
\[
\om(\ba) \equiv \int dA \Psi^*_{CS}(A) \hat{\ba} \Psi_{CS}(A) 
\]
and we make the following comments:
\begin{enumerate}
\item The above integral should be interpreted as giving the
expectation value of the algebra element $\ba$ in the state
$\Psi_{CS}$. Note that this differs from eq.~(\ref{CS-transform})
by the inclusion of the complex conjugate term. This
should not be confused with the calculation of vacuum expectation
values evaluated in Chern-Simons \emph{theory}, which is 3-dimensional
and where eq.~(\ref{CS-transform}) is interpreted as a path integral.

\item Because we need to include the complex conjugate term it is
crucial that the exponent in the Chern-Simons state is real. This
implies that we use the original complex, $SL(2,\C)$, connection
formulation of general relativity and also that this approach is only
likely to be interesting in the Lorentzian framework.

\item The above integral is of course only heuristic as we have no
suitable measure on the space of connections. Again we want to make
contact with the fact that if the algebra element $\ba$ is a
cylindrical function, i.e.\ corresponds to a configuration variable
then the above should define a diffeomorphism invariant of the graph
underlying the definition of the cylindrical function.

\item When attempting to define the above integral in this way we need
to take care that we are dealing with the non-compact gauge group
$SL(2,\C)$, which has to our knowledge not been studied in the context
of Chern-Simons knot invariants. A possible way to avoid this
difficulty by splitting the connection into real and complex parts is
suggested in~\cite{smolin95}. Note also that a rigorous treatment of the
above integrals requires both framings of the manifold $\Sigma$ and of
the graphs supporting the cylindrical functions. Progress in this
direction has been made in~\cite{major96}.

\item To complete this approach we also need to define the integral
when $\ba$ is a momentum variable, i.e.\ a triad smeared over a two
surface. To our knowledge this has not been investigated in the
literature so far. Intuitively, one expects to get a
diffeomorphism invariant of the two surfaces on which the momentum
variables are defined.

\end{enumerate}

These techniques provide a possible approach to study the
Chern-Simons state as a physical state of a well-defined theory of
quantum gravity.
The intimate relation between loop quantum gravity, knot theory and
Chern-Simons theory gives strong support for the study of this
state.

\section{Conclusions}

In this report we have focussed on three main themes. Firstly, we have
motivated why the study of the asymptotically flat sector of quantum
general relativity is important and should be pursued actively at the
present stage in the non-perturbative quantum gravity programme.
Restricting our attention to asymptotically flat space allows us to
avoid many conceptual problems facing quantum gravity while at the same
time enabling the study of a large number of physical applications. In short, the quantisation of general
relativity is most likely to succeed and produce meaningful results in
the asymptotically flat sector.

We then looked at how the standard framework of loop quantum gravity
can be extended to the asymptotically flat sector or more generally, to
the case where the spatial slice $\Sigma$ is non-compact. Here we
followed an analogy with thermal field theory and used the GNS
construction to provide us with new representations of the observable
algebra of loop quantum gravity. This gave rise to quantum theories
that can be interpreted as describing excitations of fiducial fixed
background states --- vacua of loop quantum gravity.

In the last part of this report we discussed three possible approaches
in constructing such background states. In particular this addresses
the issue of how classical phase space data or classical 3-geometries
can be approximated quantum mechanically. Together with the GNS
construction these backgrounds give us a very natural approach to 
study physical applications of loop quantum gravity, especially the
semi-classical limit. Hopefully, this will enable us to make
essential progress in uncovering the quantum picture of space and time
that loop quantum gravity provides us.

\bibliography{references}

\begin{thebibliography}{10}

\bibitem{arnsdorf99b}
M.~Arnsdorf and S.~Gupta.
\newblock Loop quantum gravity on non-compact spaces.
\newblock {\em Nucl.\ Phys.}, B577(3):529--546, 2000, e-print: gr-qc/9909053.

\bibitem{ashtekar86}
A.~Ashtekar.
\newblock New variables for classical and quantum gravity.
\newblock {\em Phys. Rev. Lett.}, 57:2244, 1986.

\bibitem{ashtekar98}
A.~Ashtekar, A.~Corichi, and J.A. Zapata.
\newblock Quantum theory of geometry \mbox{III}: Non-commutativity of
  riemannian structures.
\newblock 1998, e-print: gr-qc/9806041.

\bibitem{ashtekar95}
A.~Ashtekar, J.~Lewandowski, D.~Marolf, J.~Mour{\~a}o, and T.~Thiemann.
\newblock Quantization of diffeomorphism invariant theories of connections with
  local degrees of freedom.
\newblock {\em J. Math. Phys.}, 36:6456, April 1995, e-print: gr-qc/9504018.

\bibitem{ashtekar92}
A.~Ashtekar, C.~Rovelli, and L.~Smolin.
\newblock Weaving a classical geometry with quantum threads.
\newblock {\em Phys. Rev. Lett.}, 69:237, 1992.

\bibitem{ashtekar93}
Abhay Ashtekar and Jerzy Lewandowski.
\newblock Representation theory of analytic holonomy $c^*$ algebras.
\newblock In J.~Baez, editor, {\em Knots and Quantum Gravity}, Oxford, 1994.
  Oxford University Press, e-print: gr-qc/9311010.

\bibitem{bombelli00}
L.~Bombelli.
\newblock Statistical lorentzian geometry and the closeness of lorentzian
  manifolds.
\newblock e-print: gr-qc/0002053.

\bibitem{corichi00}
A.~Corichi and J.M. Reyes.
\newblock A gaussian weave for kinematical loop quantum gravity.
\newblock e-print: gr-qc/0006067.

\bibitem{haag93}
R.~Haag.
\newblock {\em Local Quantum Physics}.
\newblock Springer-Verlag, 1993.

\bibitem{kadison83}
R.V. Kadison and J.R. Ringrose.
\newblock {\em Fundamentals of the Theory of Operator Algebras}, volume~1.
\newblock Academic Press, 1983.

\bibitem{kodama90}
H.~Kodama.
\newblock Holomorphic wave function of the universe.
\newblock {\em Phys. Rev.}, D42:2548--2565, 1990.

\bibitem{landsman87}
N.P. Landsman and Ch.G. Weert.
\newblock Real and imaginary-time field theory at finite temperature and
  density.
\newblock {\em Phys. Rep.}, 145(3 \& 4):141, 1987.

\bibitem{major96}
Seth Major and Lee Smolin.
\newblock Quantum deformation of quantum gravity.
\newblock {\em Nucl. Phys.}, B473:267--290, 1996, e-print: gr-qc/9512020.

\bibitem{pullin93}
Jorge Pullin.
\newblock Knot theory and quantum gravity in loop space: A primer.
\newblock 1993, e-print: hep-th/9301028.

\bibitem{rovelli90}
C.\ Rovelli and L.\ Smolin.
\newblock Loop representation of quantum general relativity.
\newblock {\em Nucl. Phys.}, B331:80, 1990.

\bibitem{smolin95}
Lee Smolin and Chopin Soo.
\newblock The chern-simons invariant as the natural time variable for classical
  and quantum cosmology.
\newblock {\em Nucl. Phys.}, B449:289--316, 1995, e-print: gr-qc/9405015.

\bibitem{thiemann00}
T.\ Thiemann.
\newblock Gauge field coherent states \mbox{(GCS): I.} general properties.
\newblock 2000, e-print: hep-th/0005233.

\bibitem{thiemann00c}
T.~Thiemann and O.~Winkler.
\newblock Gauge field theory coherent states \mbox{(GCS): IV}. infinite tensor
  product and thermodynamical limit.
\newblock e-print: gq-qc/0005235.

\end{thebibliography}

\end{document}